# Liquid exfoliation of solvent-stabilised black phosphorus: applications beyond electronics

Damien Hanlon,[1,2] Claudia Backes,[1,2] Evie Doherty,[1,2,3] Clotilde S. Cucinotta,[1,2] Nina C. Berner,[1,3] Conor Boland,[1,2] Kangho Lee,[1,3] Peter Lynch,[1,2] Zahra Gholamvand,[1,2] Andrew Harvey,[1,2] Saifeng Zhang,[4] Kangpeng Wang,[1,2,4] Glenn Moynihan,[1,2] Anuj Pokle,[1,2,3] Quentin M. Ramasse,[5] Niall McEvoy,[1,3] Werner J Blau,[1,2] Jun Wang,[4] Stefano Sanvito,[1,2] David D. O'Regan,[1,2] Georg S. Duesberg,[1,3] Valeria Nicolosi,[1,2,3] and Jonathan N. Coleman,[1,2*]

[1] School of Physics, Trinity College Dublin, Dublin 2, Ireland.

[2] CRANN and AMBER Research Centres, Trinity College Dublin, Dublin 2, Ireland.

[3] School of Chemistry, Trinity College Dublin, Dublin 2, Ireland.

[4] Key Laboratory of Materials for High-Power Laser, Shanghai Institute of Optics and Fine Mechanics, Chinese Academy of Sciences, Shanghai 201800, China

[5] SuperSTEM Laboratory, STFC Daresbury Campus, Daresbury, WA4 4AD, United Kingdom

*colemaj@tcd.ie

**Abstract**

Few layer black phosphorus is a new two-dimensional material which is of great interest for applications, mainly in electronics. However, its lack of stability severely limits our ability to synthesise and process this material. Here we demonstrate that high-quality, few-layer black phosphorus nanosheets can be produced in large quantities by liquid phase exfoliation in the solvent *N*-cyclohexyl-2-pyrrolidone (CHP). We can control nanosheet dimensions and have developed metrics to estimate both nanosheet size and thickness spectroscopically. When exfoliated in CHP, the nanosheets are remarkably stable unless water is intentionally introduced. Computational studies show the degradation to occur by reaction with water molecules only at the nanosheet edge, leading to the removal of phosphorus atoms and the formation of phosphine and phosphorous acid. We demonstrate that liquid exfoliated black phosphorus nanosheets are potentially useful in a range of applications from optical switches to gas sensors to fillers for composite reinforcement.

Introduction

Over the last few years the study of two-dimensional materials has become one of the most exciting areas of nano-science. While the focus has been on the study of graphene,[1,2] more recently attention has been drawn to other two-dimensional materials such as BN, transition metal oxides and transition metal dichalcogenides such as $MoS_2$ and $WSe_2$.[3-5] These materials are of great interest both for basic research and because of their applications potential.

In the past year a new two-dimensional material has been generating considerable excitement in the research community.[6] Phosphorene is an atomically thin, 2-dimensional allotrope of phosphorus. Phosphorene monolayers (see figure 1A for structure) stack together via van der Waals interactions to form layered crystals of black phosphorus in much the same way as graphene stacks together to form graphite. Recently, it was shown that black phosphorus can be exfoliated by mechanical cleavage to form mono- and few-layer phosphorene, which we refer to as FL-BP.[7-11] This new material is only the second elemental 2D material to be studied after graphene. However, more importantly black phosphorus is a direct bandgap semiconductor in mono-, few-layer and bulk forms. The direct bandgap varies with nanosheet thickness from ~1.5 eV for monolayer phosphorene to ~0.3 eV for bulk black phosphorus.[7,12,13] This is in contrast to graphene[1] which has no bandgap and TMDs such as $MoS_2$ which display direct bandgaps only in the monolayer form.[5] Likewise Arsenene and Antimonene are expected to be semi-metals or indirect semiconductors depending on nanosheet thickness.[14] As a result black phosphorus is extremely attractive- both for electronics and optoelectronics. Mechanically cleaved phosphorus has therefore been extensively studied in applications such as transistors,[6,8,10,15-19] photodetectors[20-23] and solar cells.[8]

In addition, like other 2D materials it is likely that black phosphorus has potential to perform in a range of applications beyond (opto)electronics. Indeed, BP has already been shown to be a useful material for electrodes in lithium ion batteries.[24] Furthermore, theory predicts that FL-BP shows potential for gas sensing[25] and in thermoelectric devices.[26,27] For most applications, it will be necessary to produce black phosphorus in quantities which are orders of magnitude larger than can be achieved by mechanical exfoliation. One way to prepare nanosheets in large quantities is by liquid phase exfoliation (LPE).[28] This technique involves the sonication[29,30] or shearing[31,32] of layered crystals in appropriate liquids and has previously been applied to graphene and boron nitride as well as a range of transition metal dichalchogenides (TMDs) and oxides (TMOs).[28-30,33-36]

While phosphorene nanosheets have very recently been produced by liquid exfoliation,[37] this method remains problematic, largely because phosphorene is known to be unstable,[6,7,38] degrading via reactions with either water or oxygen. For this method to be useful, ways must be found to stabilise liquid exfoliated FL-BP nanosheets against oxidation. It is known that black phosphorus can be protected from reacting with environmental species by encapsulation, suggesting a possible way forward.[17,38] We hypothesised that liquid phase exfoliation of black phosphorus may be possible if the solvent is carefully chosen to minimise oxidation of the exfoliated nanosheets in the liquid phase due to the solvation shell acting as a barrier to prevent oxidative species reaching the nanosheet surface.

If this could be achieved it would yield numerous advantages. LPE is a powerful technique to produce nanosheets in very large quantities. In addition, the nanosheets are produced directly in the liquid phase and are thus inherently processable and can be easily formed into composites, coatings for films. This will facilitate the use of black phosphorus in a range of applications. Furthermore, control over size can easily be achieved by controlled centrifugation. This not only enables processing for applications, but is also useful for fundamental studies. For example, studying the oxidation of black phosphorus nanosheets of varying sizes would give insight into whether the degradation mechanism is associated with nanosheet edges or the basal plane.

In this work we demonstrate that black phosphorus can be exfoliated to give large quantities of few layer phosphorene nanosheets by sonication in the solvent *N*-cyclohexyl-2-pyrrolidone (CHP). We show that these nanosheets can be readily size-selected and develop quantitative spectroscopic metrics which allow the in situ measurement of nanosheet size from optical extinction spectra of liquid dispersions. The black phosphorus nanosheets are relatively stable in CHP, but degrade rapidly once water is added. In addition, when exposed to ambient conditions, black phosphorus nanosheets degrade by reaction with water predominantly at nanosheet edges. Finally we demonstrate that liquid-exfoliated black phosphorus nanosheets can be used in applications beyond electronics. We show that dispersions of black phosphorus nanosheets have impressive nonlinear optical properties allowing them to be used in optical switching applications. In addition FL-BP nanosheets can be incorporated into polymer matrices resulting in reinforced composites and are highly potent gas sensors.

**Results and Discussion**

*Exfoliation and basic characterisation*

In order to produce large quantities of black phosphorus nanosheets (see figure 1A for structure), we use liquid phase exfoliation.[29,30] This technique is often carried out in amide solvents such as *N*-methyl-2-pyrrolidone (NMP) or *N*-cyclohexyl-2-pyrrolidone (CHP), although isopropanol (IPA) has proven useful in some cases.[29] In this work we focus on the solvent CHP with IPA used to a lesser extent. Sonication of ground BP crystals (see fig 1B for scanning electron microscopic, SEM, image) in CHP yields a brown dispersion (fig 1C). In the simplest case, we remove unexfoliated material by centrifugation at 1,000 rpm (106 $g$) for 180 min to yield a stable dispersion. The dispersion obtained in this way was denoted as the standard sample (std-BP) and was subjected to further characterisation.

The successful exfoliation of BP in CHP was confirmed by transmission electron microscopy (TEM, figure 1D-F and S2.2). These images show electron-transparent, and so apparently thin, 2D nanosheets of varying lateral dimensions from <100 nm to a few microns. As shown by scanning transmission electron microscopy (STEM) and high-angle annular dark field (HAADF), the lattice is completely intact over wide regions (Figure 1 G-H). This means that BP can be exfoliated in liquids without the introduction of defects as previously observed for other materials.[28-31,39] We note that we used FL-BP exfoliated in isopropanol (IPA) for STEM and HAADF, as we found it difficult to remove the high boiling point solvent CHP completely from the nanosheets. However, qualitatively, the same intact lattice was observed from FL-BP exfoliated in CHP. To assess the lateral dimensions of the std-BP, we performed statistical TEM analysis. Depicted in Figure 1I is a histogram of nanosheet length, L, which clearly shows a bimodal size distribution with modes associated with L~100 nm and L~3 μm. We note that the nanosheets associated with the latter mode are considerably larger than is typically observed for TMDs and TMOs (e.g. L< 1 μm). [29,33,34,39-41]

To further characterise the std-BP dispersion, we measured the extinction coefficient, ε, as shown in figure 1J. This parameter is defined via the optical transmittance; $T = 10^{-\varepsilon Cl}$, where l is the cell length and C is the nanosheet concentration (measured by filtration and weighing). The extinction coefficient includes contributions from both absorbance (α) and scattering (σ),[39,42] which can be isolated using an integrating sphere and are shown in figure 1J (NB: coefficients of extinction, absorbance and scattering are related by $\varepsilon(\lambda) = \alpha(\lambda) + \sigma(\lambda)$).[39,42] Because of size dependent scattering contributions, the extinction coefficient cannot be used to accurately measure nanosheet concentration.[39,41,42] Instead, we can use the measured absorbance coefficient [α(λ=465nm) = 15 g$^{-1}$L$^{-1}$cm$^{-1}$] to give the concentration of all subsequent FL-BP dispersions. For example, figure 1K shows the black phosphorus concentration measured this way to scale with sonication time as $C \propto t^{0.4}$, similar to previous observations for graphene exfoliation.[31,43] Concentrations as high as ~1 g L$^{-1}$ can easily be realised.

To further confirm the structural integrity of the FL-BP, the dispersion was filtered onto alumina membranes and subjected to Raman and X-ray photoelectron spectroscopy (XPS). SEM confirmed the homogeneity of the film (Figure 1L inset). The Raman spectrum (Fig 1 L) shows the characteristic $A^1_g$, $B_{2g}$ and $A^2_g$ phonons of FL-BP,[7,11,12,44,45] confirming that the nanosheets are indeed black phosphorus. The P2p XPS core level spectra (Figure 1M) show the expected contributions from $P2p_{1/2}$ and $P2p_{3/2}$ components. Contributions from $P_xO_y$ species, presumably as a result of partial degradation (discussed below) are minor (<15%).

We can obtain information about the nanosheet thickness using statistical atomic force microscopy (AFM) analysis (representative images Figure 2A-C and S2.4). To overcome the problems associated with deposition of nanosheets onto substrates from high-boiling point solvents, the std-BP in CHP was transferred to IPA prior to drop casting the samples (see methods). Apparent AFM heights from liquid exfoliated nanomaterials are usually overestimated due to residual solvent.[31,39,46] To convert the apparent thickness to the number of layers, we have therefore applied a similar approach as reported for graphene and $MoS_2$[26,27] which involves the height analysis of steps associated with terraces of incompletely exfoliated nanosheets. A height profile of such a nanosheet (inset) is shown in figure 2D. By plotting the step height in ascending order (Figure 2E), it is clear that the step height is always a multiple of 2 nm. We can use this information to convert the measured apparent AFM height into number of layers, N. We can thus determine the mean number of layers of std-BP to be <N>=9.7 and find that 70% of observed nanosheets have N≤10 (Figure 2G). Shown in figure 2F is a plot of N versus nanosheet area for 150 individual flakes. This demonstrates a rough correlation between flake thickness and area: $N \propto \sqrt{A}$, which has previously been observed for exfoliated oxide nanosheets.[34]

To gain further insights in the spectroscopic properties of our liquid exfoliated FL-BP, we have selected a sample area with nanosheets of varying sizes and thicknesses by AFM (Figure 1H) and relocated the same area under a Raman microscope. A spatial Raman map ($A^1_g$ intensity, excitation wavelength 633 nm) of the region is shown in figure 1I. Single spectra (normalised to the $A^1_g$ mode) extracted at the positions indicated in the figure are displayed in figure 1J. These roughly correspond to the nanosheets circled in figure 1I. In addition, we can use our comparison to the AFM image to test, whether we observe a correlation of relative phonon intensities to varying apparent thicknesses. Intensity ratios of the respective FL-BP modes/Silicon (the substrate) confirm the validity of the approach, as the intensity ratio of the FL-BP phonon/Si increases with increasing thickness as expected (despite some scatter due to some nanosheets being smaller than the laser spot). Interestingly, we also find that the intensity ratio of the $A^1_g/B_{2g}$ phonons increases with increasing layer number (Figure 1K) for N<15. This suggests that the phonon intensity ratio can be used to estimate the layer number in case of FL-BP: $N \sim 10 A^1_g / B_{2g} - 4$.

*Size selection*

A great advantage of liquid exfoliation is that the resultant stock dispersions can be readily size-selected to yield small or large nanosheets.[31,39,47] This is important, as some applications (such as mechanical reinforcement)[48] require large nanosheets, while others (such as catalysis)[49] benefit from small nanosheets. To demonstrate that established size-selection techniques can be applied to liquid-exfoliated FL-BP, we have performed controlled centrifugation to obtain dispersions with varying size distributions (methods). Nanosheet length histograms are displayed in figure 3A-C for three representative size distributions with mean L ranging from 190 nm to 620 nm. Note that, for much of the remainder of the study, in addition of the std-BP, we have studied dispersions containing very small (S-BP, <L> = 130 nm) and very large (L-BP, <L> = 2.3 µm) to gain insights into size effects (see methods).

We have also analysed the UV-visible optical response of dispersions of nanosheets of varying size. For each sample, we have measured the optical extinction spectra as shown in figure 3D. The absorbance and scattering spectra of the size-selected FL-BP dispersions are also shown in figure 3E-F. Very clear spectral changes as a function of size can be seen in all spectra. Similar spectral changes have previously been observed for size-selected TMDs and were attributed to differences in the electronic structure associated with the edges of the nanosheets compared to the centre.[39] It is not surprising that this should be the case here as black phosphorus nanosheets are known to display edge effects.[50] We can quantify the relationship between these spectral changes and nanosheet length by plotting the ratio of optical extinction coefficient at two wavelengths (600 nm and 340 nm), $\varepsilon_{600nm} / \varepsilon_{340nm}$, *versus* nanosheet length, <L>, as shown in figure 3G. It is clear that this parameter depends on nanosheet length in a well-defined way. The scattering spectra shown in figure 3F also vary strongly with nanosheet length. Previous work has shown that, in the high wavelength regime where the optical absorbance is low, the scattering exponent scales as a power law with wavelength: $\sigma \propto \lambda^{-n}$, where n is the scattering exponent which is known to be sensitive to nanosheet length.[29,39] We estimate *n* from the data in figure 3F by fitting the curves in the range 700-800 nm. The resultant values of *n* are plotted versus inverse nanosheet length in figure 3H. Again we observe a well-defined correlation.

The data in figure 3G and H can be used to generate metrics which allow the estimation of <L> from the optical spectra. For example, by fitting an empirical function to the data in figure 2G we can show that the nanosheet length is approximately given by $L(\mu m) \approx 0.15 \ln[1.31/(0.9 - \varepsilon_{600nm}/\varepsilon_{340nm})]$. Similarly, using data in figure 3H, we can show that $L(\mu m) \approx 0.42/n$. These expressions are extremely useful because they allow the estimation of the size of liquid dispersed black phosphorus nanosheets using optical spectroscopy without any need for microscopy.

*Degradation*

It has previously been shown that exfoliated black phosphorus degrades under ambient conditions, reacting in the presence of water or oxygen, greatly limiting its application potential.[7,11,38] At first glance it might appear that this process would be accelerated for liquid exfoliated black phosphorus. However, it is entirely possible that the oxidation process is actually suppressed in certain solvents, perhaps by the solvation shell of solvent molecules protecting the nanosheets from reacting with water. This makes it important to monitor the temporal stability of liquid dispersed black phosphorus. To do this, we have tracked the absorbance, $A(\lambda=465nm)$, of std-BP, S-BP and L-BP as a function of time (Figure 4A, N.B $A = \alpha C l$). Degradation of the FL-BP should result in a fall in measured absorbance over time, which we do indeed observe in all samples. However, it is clear that the larger flakes are more stable than those in the standard sample which are in turn more stable than the small flakes, suggesting that the chemical reaction starts from the edge as previously found for other 2D materials such as $TiS_2$.[51] We expect this degradation to follow a complex, multistep reaction (see below and SI). Thus, we fit these curves to an empirical function of the form: $A = A_{stable} + A_{unstable}e^{-t/\tau}$, where $A_{unstable}$ represents the total amount of FL-BP which degrades over time. For the data in fig 4A, the time constant falls from τ=350 to 190 to 115 hrs as the flake size increases from the S-BP (<L>=130 nm) to std-BP (<L>=1000 nm) to L-BP (<L>=2300 nm) samples. Interestingly, $A_{stable}/(A_{stable}+A_{unstable}) \approx 75\%$, independent of nanosheet size, suggesting the degradation to be limited by the amount of reactant present (below we determine this reactant to be water). Importantly, after 3 days, only ~8 % of the std-BP had dissolved (as confirmed by XPS), showing these systems to be stable enough to allow processing for applications. We note that the degradation is significantly faster in other solvents such as IPA (Figs. S4.3-4.5), strongly suggesting that the FL-BP is protected from the environment by factors specific to CHP (or potentially other pyrrolidone-based solvents).

Since it has been proposed that the adsorption of and reaction with water drives the degradation of the FL-BP, we have also investigated the degradation following addition of water to the dispersion. Adding small amounts of water significantly accelerates the degradation relative to pure CHP (Fig 4B). As the water content is increased, both τ and $A_{stable}/(A_{stable}+A_{unstable})$ are reduced to 40 hrs and 10% respectively, for the sample with 5% added water (fig S4.7). In comparison, bubbling oxygen through the dispersion had virtually no effect on the absorbance, suggesting water to be predominately responsible for the oxidation process.

Furthermore we investigated the degradation of liquid exfoliated FL-BP after deposition onto substrates using AFM, Raman and TEM. In these experiments, we monitored the same sample region (*i.e.* the same nanosheets) with AFM and Raman from immediately after deposition until 11 days later. As shown by the AFM images in figure 4C (and Figs. S4.8-4.11), we observe broadening and blurring (*i.e.* loss of fine structure) of the nanosheets (height profile figure 4D and S4.11). This is typically accompanied with shrinking in lateral dimensions. In some cases, we observe an increased apparent height which is typically accompanied with an increased contrast in the phase images, attributed to the adsorption of water. However, it is important to note, that even after 11 days of ageing under ambient conditions, the liquid exfoliated FL-BP from CHP is remarkably stable, as the average Raman spectra (normalised to Si) of the same sample region show no spectral changes (figure 4E and S4.13). This implies that the overall structure of the FL-BP nanosheets does not change with time, consistent with degradation occurring at nanosheet edges. This is further confirmed by TEM analysis tracking the same nanosheets over time (figure 3F and Figs. S4.14-4.15). While the flakes look disrupted and, similar to AFM, blurred out at the edges, most of the nanosheets remain intact even after 16 days. We note that changes in the apparent morphology may also partly be related to beam damage. It is important to emphasize that we clearly observe degradation when using a less favourable, low boiling point, hygroscopic solvent such as IPA. Taken together, this suggests that i) CHP indeed protects the nanosheets from degrading and ii) degradation starts at edges.

*Modelling*

In order to explain the FL-BP degradation in water, we have performed density functional theory (DFT) calculations. Since we experimentally observe a drop in pH with time after water addition and the formation of phosphates and phosphonates (XPS), we propose the following degradation reaction.

$$BP + 3H_2O \rightarrow BP_{2VAC} + PH_3 + H_3PO_3 \qquad (1)$$

This reaction was evaluated for both edge sites and basal plane P atoms. In both cases, a defective structure $BP_{2VAC}$, with two P vacancies in the BP supercell is formed (2.5% vacancy concentration). The reaction energy of process (1) is evaluated as

$$\Delta E=[E(BP)+ 3E(H_2O)] - [ E(BP_{2VAC})+ E(PH_3)+ E(H_3PO_3)] \quad (2)$$

where $[E(BP)+ 3E(H_2O)]$ is the energy of the nanosheet and three isolated water molecules at infinite distance from the nanosheet and $[E(BP_{2VAC})+ E(PH_3)+ E(H_3PO_3)]$ is the energy of the defective flake infinitely distant from the other isolated reaction products. When occurring at the edge, this reaction is exothermic ($\Delta E$= -1.2 eV, see Figure 4). In contrast, the reaction is endothermic ($\Delta E$= 0.258 eV) when the process occurs far from the edge. However, whether occurring at the edge or in the middle of BP, degradation is a multistep process. Since we are interested in whether a reaction starts at the edge or the basal plane, we have analysed in more detail the early steps in the reaction (approach of $H_2O$ and splitting of $H_2O$ to hydroxyl group and H atoms chemisorbed to neighbouring P). In both edge and basal plane case, the water adsorption is slightly exothermic, while the splitting of the water is highly unfavourable on the basal plane. We have furthermore ruled out that oxidation could proceed in the middle of BP nanosheet if activated by the formation of the first hole by studying the degradation process starting from a defective structure $BP_{2VAC}$. The reaction is again endothermic by approximately $\Delta E$=0.26 eV further supporting that the reaction is unlikely to occur on the basal plane.

*Applications*

Even though liquid-exfoliated FL-BP does degrade under ambient conditions, the degradation timescale is slow enough to allow processing of nanosheets for applications testing in a number of areas. We believe that if applications potential is demonstrated, practical usage of FL-BP will be enabled by encapsulation.[17,38] In this work we have chosen three applications which have not previously been described to demonstrate the broad potential of FL-BP.

The nonlinear optical response of the FL-BP dispersions was investigated by open-aperture Z-scan system in conjunction with a 340 fs pulsed fiber laser.[52] In Figure 6 A-B, it is clearly seen that normalized transmission of the FL-BP dispersions increases with laser intensity at both 1030 nm and 515 nm. Such broadband saturable absorption (SA) suggests that the FL-BP nanosheets could serve as ultrafast nonlinear saturable absorber, an essential mode-locking element for ultrashort pulsed lasers.[53,54] Since graphene is a well-known broadband saturable absorber,[55] we carried out the same nonlinear measurement for graphene dispersions prepared by the similar liquid exfoliation. As shown in Figure 6A-B, at the same level of linear transmission, the FL-BP dispersions exhibit much stronger SA response than the graphene dispersions at both wavelengths. The saturable intensity $I_s$ is obtained by fitting the Z-scan data with the SA model $dI/dz = -\alpha I$ where $\alpha = \alpha_0/(1+I/I_s)$: $\alpha_0$ is the linear absorption coefficient and $I$ is the excitation intensity. As shown in Figure 6C, $I_s$ of FL-BP is much lower than that of graphene at both 1030 nm and 515 nm when the linear transmission is equal. The significant ultrafast nonlinear property of FL-BP implies a huge potential in the development of nanophotonic devices, such as mode-lockers, Q-switchers, optical switches, etc.[56]

Theoretical work has suggested FL-BP nanosheets as gas sensors.[25] We prepared thin films of FL-BL nanosheets by vacuum filtration followed by transfer to interdigitated electrode arrays. Two-probe measurements showed relatively high conductivities of ~1 S/m, similar to films of WTe$_2$ nanosheets.[57] Shown in figure 6D is the resistance change of a thin film of FL-BP nanosheets on exposure to ammonia (NH$_3$) gas. We observe a resistance increase, consistent with NH$_3$ donating electrons to the p-type FL-BP. By extrapolation of the signal-to noise level (figure 6E), and assuming the minimum detectable signal to be 3× the RMS noise level, we estimate a detection threshold of 80 ppb. This shows FL-BP networks to be competitive with other nano-sensors[58-60] and a very promising material for gas detection.

The impressive mechanical properties of 2D materials in general[61] suggest the potential to use FL-BP as a reinforcing filler in composites. Shown in figure 6F are the representative stress strain curves for a film of polyvinylchloride (PVC) and a PVC:FL-BP (0.3 vol%, see methods). It is clear that that the mechanical properties improve considerably both in the high and low strain regimes. Shown in figure 6 G-I are the composite modulus, strength and tensile toughness, plotted as a function of black phosphorus loading content. In each case, mechanical properties increase considerably for loading levels of only 0.3 vol%. The modulus, Y, increases from 500 MPa for PVC to 900 MP for the 0.3 vol% composite. In addition, at 0.3 vol% loading, the measured strength of the composite doubles while its tensile toughness displays a six-fold increase. These are significant increases and are competitive with those found using graphene as a filler in both PVC[62] and polymers in general.[63]

We used first-principles simulations (see methods) to obtain the bulk black phosphorus Young's modulus as a function of orientation in the plane of the crystal as shown in figure 6 J, yielding a Hill in-plane average modulus of $\langle Y_{BP} \rangle = 97$ GPa. We also calculated the Poisson's ratio finding it to be highly anisotropic. Assuming the BP nanosheets lie in-plane,[64] we can use the calculated $\langle Y_{BP} \rangle$, coupled with the rule of mixtures[63] to predict the composite modulus as a function of BP volume fraction, V$_f$: $Y_{comp} = \langle Y_{BP} \rangle V_f + Y_{poly}(1-V_f)$. We find this prediction (blue line) to agree well with the experimental data even though we have not corrected for the finite aspect ratio of the nanosheets or their layered nature.

**Conclusion**

In conclusion, we have shown the black phosphorus crystals can be efficiently exfoliated in appropriate solvents to yield high quality, few-layered nanosheets with controllable size. The nanosheets are remarkably stable in CHP, probably due to protection by the solvation shell. However, removal of CHP or addition of water results in degradation of the nanosheets. Both experimental and computational studies indicate the degradation to occur at the nanosheet edge and proceed by reaction of the FL-BP with water yielding phosphine and phosphorous acid as by products. We demonstrate that liquid-exfoliated FL-BP nanosheets have potential for use in applications as saturable absorbers, gas sensors and reinforcing fillers for composites. We believe this work is important as it will facilitate the production and use of FL-BP in a broad range of applications.

**Methods**

Additional information on materials and methods are provided in the SI.

*Sample preparation*

Black phosphorus crystals were purchased from Smart Elements (purity 99.998%) all other materials from Sigma Aldrich and used as received. BP was lightly ground with pestle and mortar and immersed in CHP (concentration 2 g/L). The dispersion was sonicated for 5 hours at 60% amplitude with a horn-probe sonic tip (VibraCell CVX, 750W) under cooling yielding a stock disperison. During sonication a sample (4ml) was taken at specific time periods over the course of ~80h. These samples were subsequently centrifuged at 1krpm for 180 min and subjected to absorbance spectroscopy. Aliquots of the stock dispersion were centrifuged at 1,000 rpm (106 *g*) for time periods varying from 5 to 240 min in a Hettich Mikro 220R centrifuge equipped with a fixed-angle rotor 1016. The supernatant was decanted and subjected to absorbance spectroscopy. The supernatant with centrifugation conditions 1,000 rpm for 180 min was denoted standard sample (std-BP). The std-BP was subsequently separated into small and large stable nanosheets. For this purpose, aliquots of std-BP dispersion were subjected to an additional centrifugation of 5 krpm (2,660 *g*) for 120 min. The supernatant (containing small flakes) was decanted and characterised as S-BP while the sediment (containing large flakes) was re-dispersed in fresh CHP and characterised as L-BP. Alternatively, the FL-BP was size selected by controlled centrifugation with subsequently increasing rotation speeds. The sediment after 2 krpm (426 $g$, 2 h) was discarded, while the supernatant was subjected to further centrifugation at 3 krpm (958 $g$, 2 h). The sediment was collected in fresh solvent, while the supernatant was subjected to further centrifugation at 4 krpm (1,702 $g$, 2 h). Again, the sediment was collected and the supernatant centrifuged at high rpm. This procedure was repeated for 5 krpm (2,660 $g$, 2 h), 10 krpm (10,170 $g$, 2 h) and 16 krpm (25,000 $g$, 2 h) to yield samples with decreasing sizes in the respective sediments.

*Characterisation:*

Optical extinction and absorbance was measured on a Perkin Elmer 650 spectrometer in quartz cuvettes. To distinguish between contributions from scattering and absorbance to the extinction spectra, dispersions were measured in an integrating sphere using a home-built sample holder to place the cuvette in the centre of the sphere. The absorbance spectrum is obtained from the measurement inside the sphere. A second measurement on each dispersion was performed outside the sphere to obtain the extinction spectrum. This allows for the calculation of the scattering spectrum (extinction-absorbance).

Bright field transmission electron microscopy imaging was performed using a JEOL 2100, operated at 200 kV while HRTEM was conducted on a FEI Titan TEM (300 kV). High resolution TEM images (Figure 4F) were taken using an FEI Titan 60-300 Ultimate Microscope operated at 300 kV. The FL-BP was dropped on grids using a drop casting method and excess fluid was absorbed by an underlying filter membrane. It was then baked in vacuum at $120^0$C for several hours. The samples were imaged on the day they were received which is termed Day 1. The same flake was then imaged on Day 3 and 16 respectively. No changes in nanosheet structure of morphology were observed between Day 1 and Day 3, but by Day 16 a combination of reaction products and water adsorption results in a liquid layer on the flake. This layer can be removed with the beam and comparison of the shape of the flakes between Day 1 and Day 16 shows that the overall shape and size of the flake has not changed. There is also lattice apparent in the flake on Day 16 when using a high magnification (×300k).

Aberration-corrected scanning transmission microscopy (STEM) images were taken using a Nion Ultrasteme 100 (cold filed emission gun (FEG) at the SuperSTEM Laboratory in Daresbury, UK. The suspended FL-BP flakes were dropped onto lacey carbon coated copper TEM grids as before. The samples were then prebaked at $120^0$C in a vacuum overnight. The images were recorded using a 100 kV acceleration voltage using a high field annular dark field detector (HAADF) and low-pass bright field imaging. Figure 1F shows a representative Butterworth-filtered HAADF STEM image and G shows a low-by-pass bright field STEM image of the material. The lattice shows high uniformity with the presence of very few defects or dislocations. Depending on the zone axis used, the lattice can exhibit a "dumbbell" type configuration.

Atomic force microscopy (AFM) was carried out on a Veeco Nanoscope-IIIa (Digital Instruments) system equipped with an E-head (13 μm scanner) in tapping mode after depositing a drop of the dispersion transferred to IPA on a pre-heated (150 °C) Si/SiO$_2$ wafer with an oxide layer of 300 nm. Raman spectroscopy on individual flakes was performed using a Horiba Jobin Yvon LabRAM HR800 with 633 nm excitation laser in air under ambient conditions. X-ray Photoelectron Spectroscopy was performed under ultra-high vacuum conditions ($<5\times10^{-10}$ mbar), using monochromated Al Kα X-rays (1486.6 eV) from an Omicron XM1000 MkII X-ray source and an Omicron EA125 energy analyser. An Omicron CN10 electron flood gun was used for charge compensation and the binding energy scale was referenced to the adventitious carbon 1s core-level at 284.8 eV. Core-level regions were recorded at an analyser pass energy of 15 eV and with slit widths of 6 mm (entry) and 3 mm × 10 mm (exit), resulting in an instrumental resolution of 0.48 eV. After subtraction of a Shirley background, the core-level spectra were fitted with Gaussian-Lorentzian line shapes and using Marquardt's algorithm.

*Modelling*

Quantum Espresso code was used to evaluate the reaction energies. In this case we selected an energy cutoff of 50 ryd, a grid of 2×2×1 Monchorst-Pack k-points (see SI for details about the supercells) and ultrasoft pseudopotentials. Langreth and Lundqvist van der Waals corrected exchange and correlation functional[65] was used and atomic forces were relaxed until they were smaller than $5\times10^{-3}$ eV/A.

Such calculations were also performed in order to clarify the microscopic origins of an observed dramatic increase in the Young's modulus of composites reinforced with BP nanosheets. In this case, converged PBE calculations augmented with the Grimme dispersion correction[66] were performed in order to compute the elastic constants of bulk phosphorus by finite-difference, the nanosheets in question being closer to the bulk than monolayer regime. The Voigt-Reuss-Hill approach,[67] was used to estimate both isotropic (found to be comparable to current literature[68]) and nanosheet in-plane-only estimates of the average nanosheet Young's modulus, as well as upper (Voigt) and lower (Reuss) expected error bounds.

*Applications*

For non-linear optical measurements, an open-aperture Z-scan system was used to study the ultrafast nonlinear optical properties of the FL-BP (std-BP) and graphene dispersions. This measures the total transmittance through a sample as a function of incident laser intensity, while the sample is sequentially moved through the focus of a lens (along the z-axis).[52,54] All experiments were performed with 340 fs pulses from a mode-locked fiber laser, which was operated at 1030 nm and its second harmonic, 515 nm, with a pulse repetition rate of 1 kHz. All dispersion samples were tested in quartz cuvettes with 1 mm pathlength.

Gas sensing was conducted on the FL-BP (std-BP) prepared in CHP and then subsequently transferred into 2-proponal (see AFM experimentals) to facilitate filtration onto a nitrocellulose membrane. Following filtration the film was allowed to dry under vacuum conditions. The FL-BP film was cut into 10×2mm rectangular pieces. These were then transferred onto silicon dioxide wafer while the nitrocellulose membrane was dissolved using the transfer method according to Wu *et al.*[69]

For gas sensing, gold electrodes were sputtered on top of an adhesion layer of nickel (Ni/Au = 30/70 nm) using a metal shadow mask, which has a 2 mm wide and 200 μm long channel. All devices were loaded in a gas sensing chamber and annealed at 100 °C for 1 hour to remove residues and adsorbates on the surface. The gas sensing chamber was kept at room temperature at pressure 10 Torr, with a 100 sccm flow of the $NH_3$ and $N_2$ mixtures. The resistance change of five devices upon interval gas exposure was simultaneously measured using a Keithley model 2612A SourceMeter and a Keithley 3706 System Switch at a constant bias voltage of 1 V. The initial resistance and RMS noise were calculated from the first 500 data points, approximately 2 min, before the first gas injection. $NH_3$ for 2 min and pure $N_2$ for 5 min were periodically introduced to record sensor response and recover, respectively.

A previously prepared nanofiller FL-BP dispersion in N-Cyclohexyl-2-pyrrolidone, centrifuged between 1krpm for 180min and 3k for 120 mins and redispersed in fresh CHP solvent, was subsequently filtered onto a Polyester Membrane Filter (0.2 micron) of known mass. The membrane was dried in a vacuum oven at 100C for 2 h and the mass of the membrane remeasured to attain the mass of the filtered nanofiller. The nanofiller was redispersed by bath sonication (Branson 1510 Model 45 kHz) in a 65:35 tetrahydrofuran (THF), chloroform solvent mixture. Poly (vinyl chloride) (PVC) was dissolved in a solvent mixture (65:35 tetrahydrofuran, chloroform). A range of FL-BP/PVC/THF/Chloroform dispersions (from 0 to 0.0074 Vf) were made by adding the FL-BP/THF/Chloroform filler solution to the PVC/THF/Chloroform solution with varying increments of loading. These solutions were of constant mass (150 mg of FL-BP & polymer) and constant volume (5 mL of FL-BP, polymer & solvent). These samples were sonicated in the same bath as before for 1 h to homogenise after the blending of the solutions. The homogenised solution mixtures were then dropcast into 5cm×5cm×1cm Teflon trays and placed in a vacuum oven for 4 hours at 40 °C under no vacuum to form composite films. The films were then kept overnight (~17 h) at 50°C under full vacuum to ensure that the solvent was completely removed and to protect the filler material from decomposition before testing. For mechanical measurements, the films were cut into 2.25mm strips and then tested on a Zwick Roell tensile tester with a 100N load cell at a strain rate of 10mm/min.

Acknowledgements

The research leading to these results has received funding from the European Union Seventh Framework Programme under grant agreement n°604391 Graphene Flagship. We have also received support from the Science Foundation Ireland (SFI) funded centre AMBER (SFI/12/RC/2278). In addition, JNC acknowledges the European Research Council (SEMANTICS) and SFI (11/PI/1087) for financial support. CB acknowledges the German research foundation DFG (BA 4856/1-1). ED, AP and VN acknowledge ERC 2DNanoCaps, SFI PIYRA and FP7 MoWSeS. GSD, KL, NMcE and NB acknowledge SFI for PI_10/IN.1/I3030. STEM experiments were performed at SuperSTEM, the EPSRC UK national facility for aberration-corrected STEM. JW and SZ acknowledge NSFC (61178007 and 61308034), STCSM (12ZR1451800) and the External Cooperation Program of BIC, CAS (No. 181231KYSB20130007). SS and CSC have been supported by the European Research Council (Quest project). All calculations were performed on the Parsons cluster maintained by the Trinity Centre for High Performance Computing, (project HPC_12_0722). This cluster was funded through grants from Science Foundation Ireland.

DH prepared liquid dispersions, DH, CB, AH, NMcE and GSD performed spectroscopy, CB performed AFM/Raman, DH performed low-res TEM, ED, AP, QMR and VN performed hi-res TEM, CBo, ZG and PL prepared and tested composites, KL and GSD tested gas sensing CSC and SS performed calculations on FL-BP stability, GM and DDO'R performed simulation of composite mechanical reinforcement by FL-BP loading, SZ, KW, WB and JW performed nonlinear optical measurements, JNC and CB planned the experiments and wrote the paper.

**Figures**

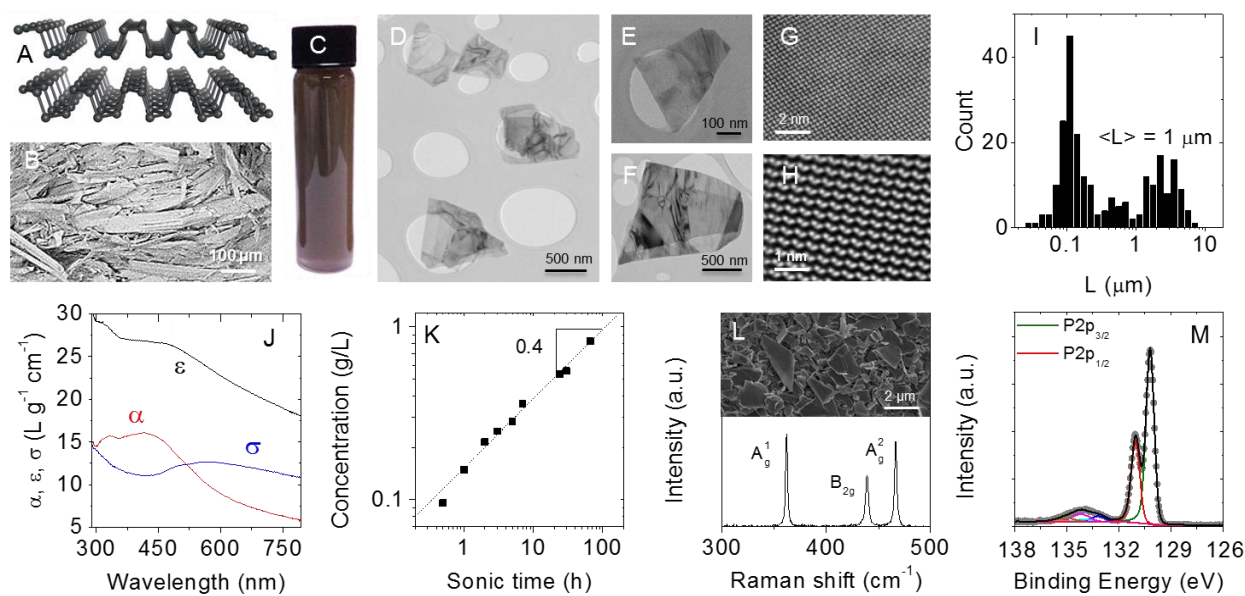

**Figure 1:** A) Structure of Black Phosphorus (BP). B) SEM image of a layered BP crystal. C) Photograph of a dispersion of exfoliated FL-BP in CHP. D-F) Representative low-resolution transmission electron microscopic (TEM) images of FL-BP exfoliated in *N*-cyclohexyl-2-pyrrolidone (CHP). G) Low-by-pass bright-field scanning transmission TEM (STEM) image and H) Butterworth filtered high-angle annular dark field (HAADF) STEM image of FL-BP (exfoliated in isopropanol) showing the intact lattice. I) Nanosheet length histogram of the exfoliated FL-BP obtained from TEM. J) Extinction, absorbance, scattering coefficient spectra of FL-BP in CHP. K) Concentration of FL-BP as a function of sonication time. The dashed line shows power law behaviour with exponent 0.4. L) Raman spectrum (mean of 100 spectra, excitation 633 nm) of a filtered dispersion. Inset: Scanning electron microscopic image of thin film. M) X-ray photoelectron spectroscopy P core level region.

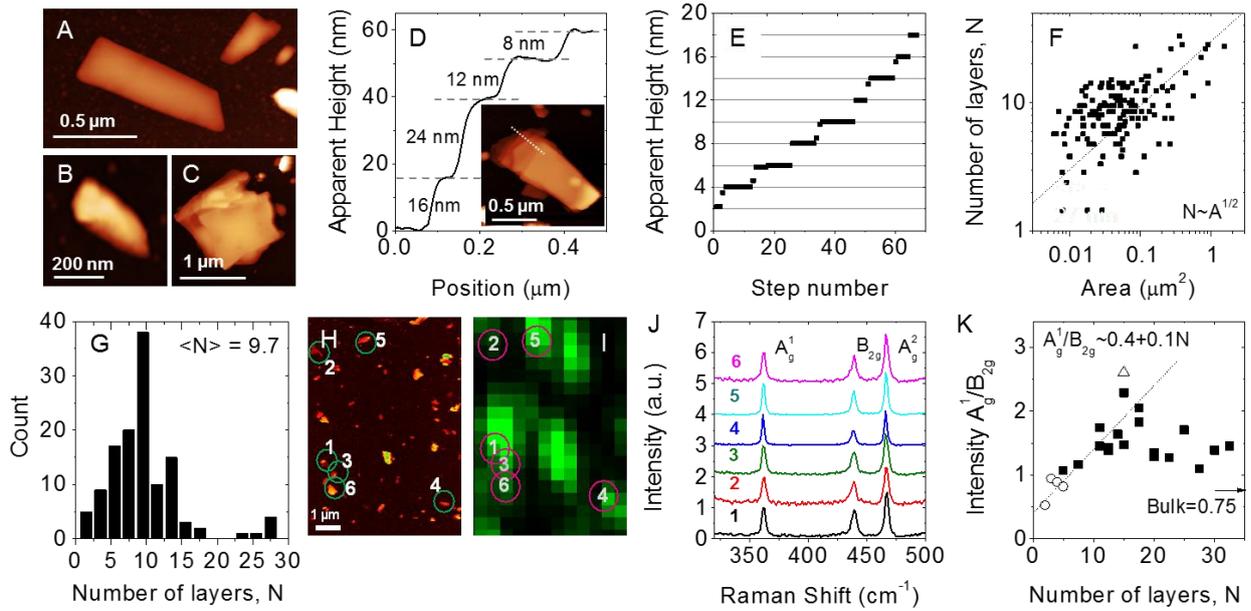

**Figure 2.** A-C) Representative atomic force microscopic (AFM) images. D) Height profile of the nanosheet in the inset along the line showing clearly resolvable steps. E) Heights of steps of deposited FL-BP nanosheets. The step height is always found to be a multiple of 2 nm. F) Plot of number of layers per nanosheet as a function of flake area determined from AFM. The dashed line indicates $N \propto \sqrt{A}$ behaviour. G) Histogram of number of monolayers per nanosheet. H-I) Large area AFM image (H) and Raman $A^1_g$ intensity map (I) of the same sample region (excitation wavelength 633 nm). J) Raman spectra (normalised to $A^2_g$) of the nanosheets indicated in H and I. K) Plot of intensity ratio of the $A^1_g$ / $B_{2g}$ phonons as a function of layer number. Both open circles (ref[70]) and triangle (ref[12]) show data extracted from published papers. The dashed lines shows an approximate linear relationship between Raman ratio and nanosheet thickness. The Raman ratio associated with bulk BP is indicated by the arrow (ref[70]),

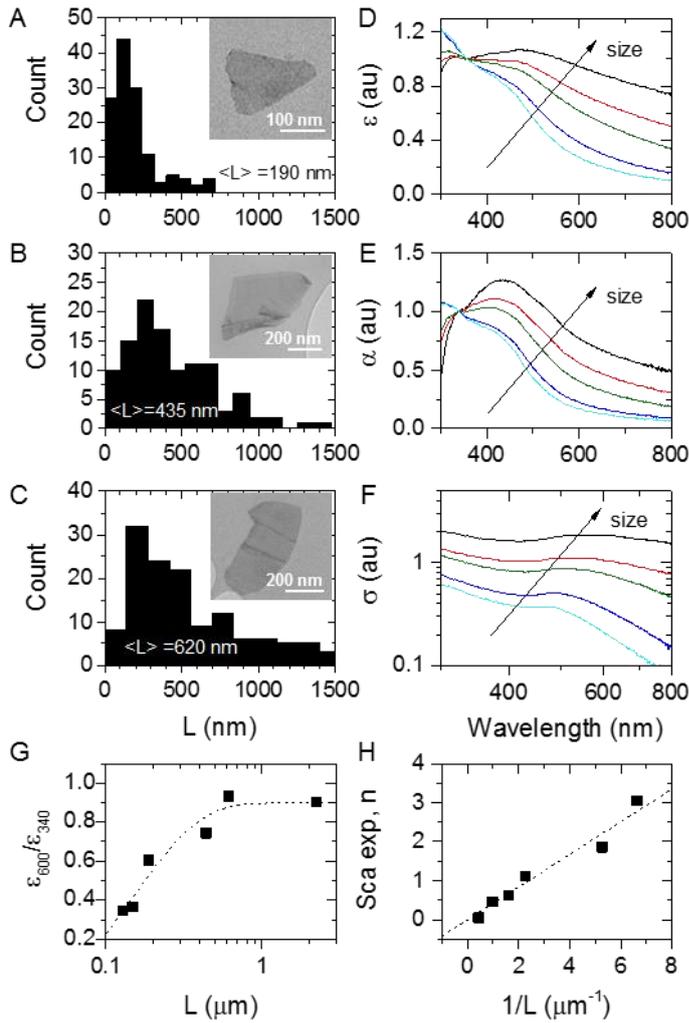

**Figure 3:** A-C) TEM length histograms of size-selected FL-BP in CHP including representative TEM images as insets. D) Extinction (ε) spectra normalised to 355 nm of FL-BP dispersions with different mean nanosheet lengths showing systematic changes as a function of size. Extinction spectra can be split into contributions from absorbance (α) and scattering (σ): E) Absorbance spectra of the same dispersions (normalised to 340 nm) and F) scattering spectra. Scattering spectra were obtained by subtracting the absorbance spectra from the normalised extinction spectra. G) Ratio of measured extinction at 600 nm to that at 340 nm plotted versus nanosheet length. H) Long wavelength scattering exponent, *n*, plotted versus inverse nanosheet length. The parameters plotted in G and H can be used as metrics to estimate mean nanosheet length from the optical spectra.

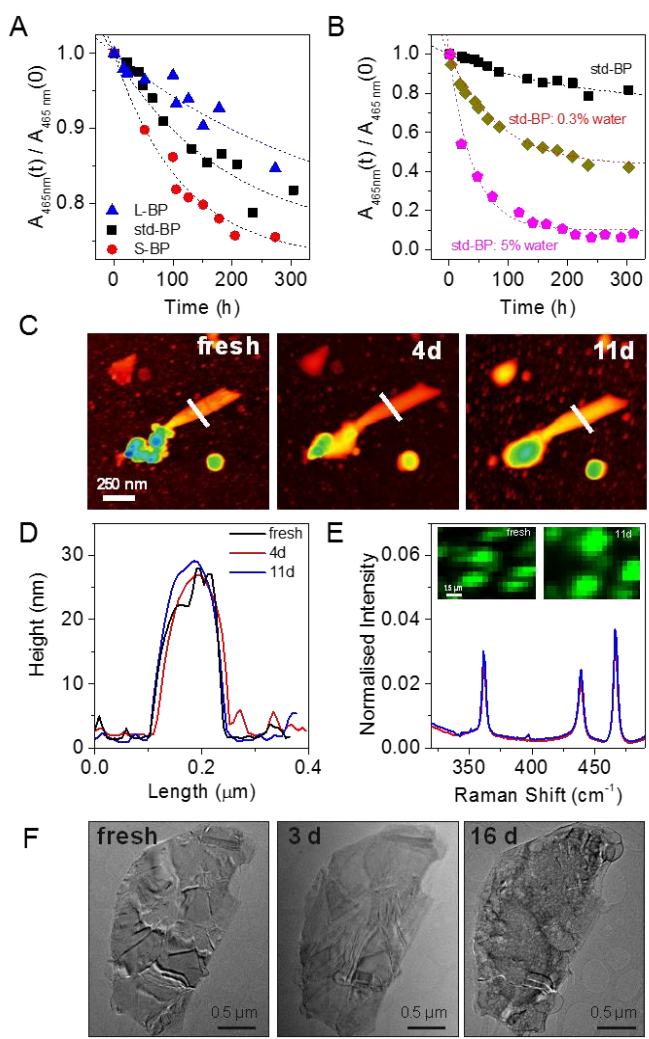

**Figure 4:** A) Relative absorbance at 465 nm, measured as a function of time, for the standard FL-BP dispersion (std-BP) and size selected dispersions containing small (S-BP) and large nanosheets (L-BP). Small nanosheets degrade significantly faster than large nanosheets. B) Relative absorbance at 465 nm, measured as a function of time, for the standard BP dispersion (std-BP) and std-BP dispersions with 0.3Vol% and 5Vol% of water in the FL-BP dispersions. In A) and B), the dashed lines represent exponential decays. C) Sequence of AFM images of the same sample region of an as-prepared sample, after four days and 11 days, respectively. D) AFM height profiles of the line in C. Fine structure of the flake is lost and edges are softened. E) Mean Raman spectra (633 nm excitation) summed over the same sample region of an as-prepared sample, after four days and 11 days, respectively. Spectra are normalised to the silicon peak at 521 nm. Inset: Raman $A^1_g$ intensity map of the sample region (freshly prepared and after 11). F) Sequence of TEM images of the same flake of a freshly prepared dispersion, after 3 days and 16 days, respectively.

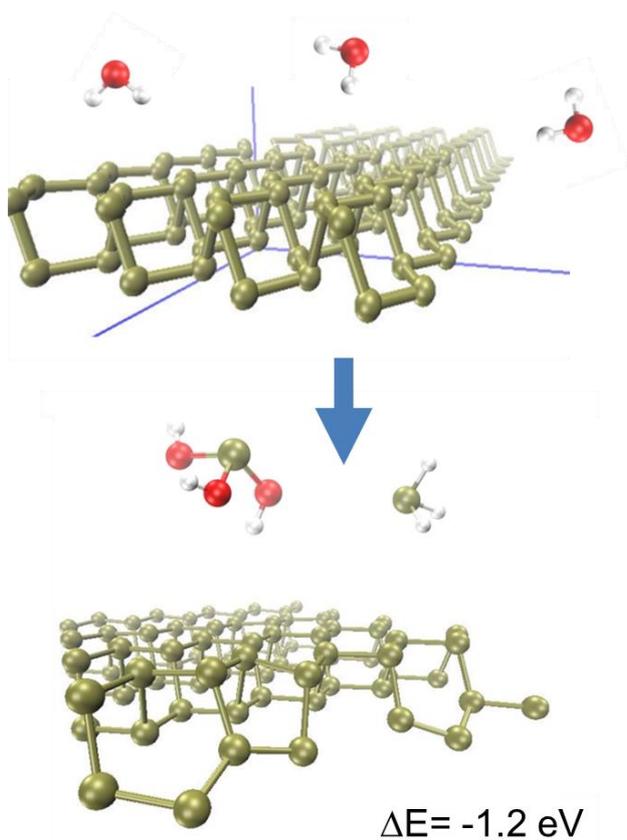

**Figure 5:** Edge selective degradation model for BP exposed to pure neutral water. Top and bottom panel represent reagent (BP edge + three water molecules) and reaction products (BP defective edge + phosphine + phosphorous acid), respectively, with the reaction energy also given. Green, red and white balls represent P, O and H atoms, respectively.

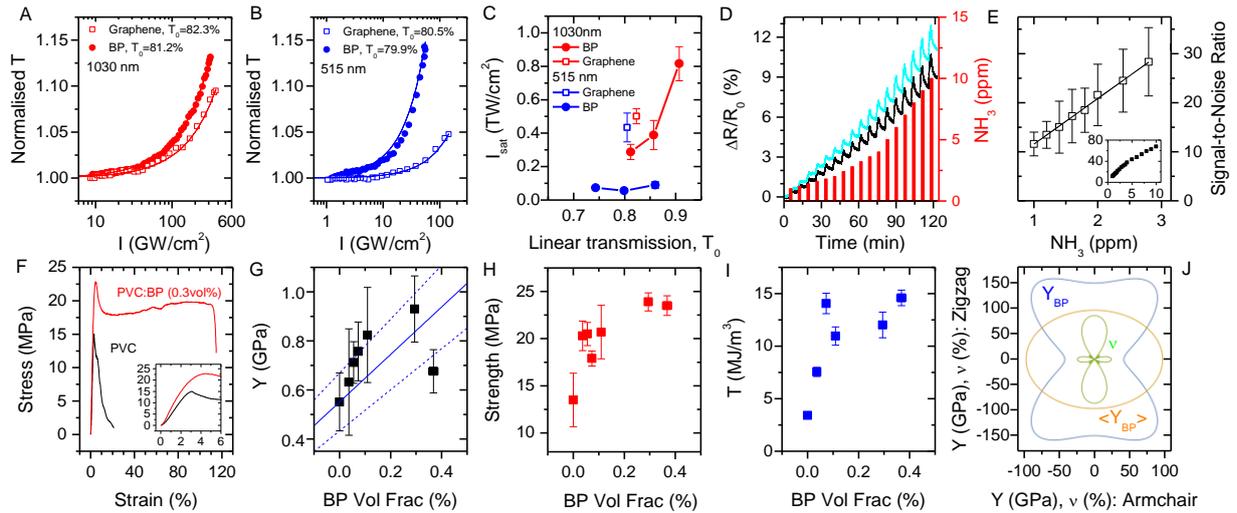

Figure 6: Applications of liquid exfoliated FL-BP. A-B) Saturable absorption of std-BP and graphene in CHP for fs pulses excited at (A) 1030 nm and (B) 515 nm. Linear transmission $T_0$ is given in the legend. C) Saturation intensity of FL-BP and graphene as a function of $T_0$. D-E) Sensing of $NH_3$ gas using std-BP films. D) Sensor response plot show percentile resistance change versus time of the FL-BP films with a bias voltage of 1 V at room temperature, upon consequent NH3 exposures at various concentrations from 1 to 10 ppm. E) Plot of signal-to-noise ratio as a function of NH3 concentration from 1 to 3 ppm (inset: from 1 to 10 ppm). The error bar represents the standard deviation of five devices and the linear line indicates the fitted line. F) Representative stress strain curves for PVC and PVC: FL-BP (0.3vol%). Inset: low strain regime. G-I) Young's modulus, including the theoretical constant-strain rule-of-mixtures modulus prediction (blue using the Hill planar-averaged nanosheet modulus, blue-dashed using its Voigt and Reuss bounds) (G), tensile strength (H) and tensile toughness (I) plotted as a function of FL-BP volume fraction. J) Calculated orientation-dependence of the bulk black phosphorus in-plane-only Young's modulus (blue), its Hill average (yellow) and Poisson's ratio (green).